\documentclass{article}
 \usepackage{natbib}
 \usepackage{bm}
\usepackage{tikz}
\usepackage{booktabs}
\usepackage{threeparttable}
\usepackage{adjustbox}
\usepackage{colortbl}
 \usepackage{amsmath}
\usepackage{hyperref}
\usepackage{longtable}
\usepackage{algorithm}
\usepackage{algorithmic}
\usepackage{multirow}
\usepackage{threeparttablex}
\usepackage{makecell}
\usepackage{dsfont}
\usepackage{tabularx}
\usepackage{subcaption} 
\usepackage{enumitem} 
\usepackage{lmodern} 
\usepackage{authblk}
\usepackage{amsfonts}
\usepackage{natbib} 
\usepackage[a4paper,left=2cm, right=2cm, top=1.50cm, bottom=1.50cm]{geometry}

\newcommand{\mat}[1]{\bm{#1}} 
\newcommand{\obs}{{\text{obs}}} 
\newcommand{\mis}{{\text{mis}}} 
\newcommand{\mar}{{\text{MAR}}} 

\hypersetup{
    colorlinks=true,
    linkcolor=blue!50!black,
    citecolor=blue!50!black,
    urlcolor=blue!50!black,
    pdfborder={0 0 0} 
}

\title{Hybrid methods for missing categorical covariates in Cox model}

\tiny{\author[1,2]{\small Abdoulaye Dioni}
\author[1,3]{Lynne Moore}
\author[1,2]{Aida Eslami}
\affil[1]{ Département de Médecine Sociale et Préventive
  Université Laval, Québec, Canada}
\affil[2]{ Institut Universitaire de Cardiologie
 et de Pneumologie de Québec, Québec, Canada}
 \affil[3]{Centre de recherche du CHU de Québec,  Québec, Canada}}
\date{}

\begin{document}
\maketitle
\vspace{2cm}

\section*{Abstract}
Survival analysis aims to explore the relationship between covariates and the time until the occurrence of an event. The Cox proportional hazards model is commonly used for right-censored data, but it is not strictly limited to this type of data. However, the presence of missing values among the covariates, particularly categorical ones, can compromise the validity of the estimates.  
To address this issue, various classical methods for handling missing data have been proposed within the Cox model framework, including parametric imputation, nonparametric imputation, and semiparametric methods. It is well-documented that none of these methods is universally ideal or optimal, making the choice of the preferred method often complex and  challenging.
To overcome these limitations, we propose hybrid methods that combine the advantages of classical methods to enhance the robustness of the analyses. Through a simulation study, we demonstrate that these hybrid methods provide increased flexibility, simplified implementation, and improved robustness compared to classical methods. The results from the simulation study highlight that hybrid methods offer increased flexibility, simplified implementation, and greater robustness compared to classical approaches. In particular, they allow for a reduction in estimation bias; however, this improvement comes at the cost of reduced precision, due to increased variability. This observation reflects a well-known methodological trade-off between bias and variance, inherent to the combination of complementary imputation strategies.

\textbf{Keywords}: Missing covariate, Cox proportional hazards model, missing at random, classical imputation methods, hybrid imputation methods.

\newpage
\section{Introduction}\label{intro:chp2}

The proportional hazards model introduced by \citet{Cox1972} is widely used to analyse right-censored survival data \citep{Tsiatis2006, Kleinbaum2012}. It is a semiparametric model that combines a parametric component (covariate coefficients) with a nonparametric component (baseline hazard function) \citep{Lawless2003, Moore2016, Tsiatis2006}. One of the main advantages of the Cox model is that the estimated coefficients directly represent hazard ratios \citep{Kleinbaum2012, Lee1997}.  
In practice, the Cox model is commonly used in the fields of social and health sciences. One of the challenges in analyzing data from these research areas is the potential presence of missing values in some explanatory covariates used as independent variables in statistical models. Therefore, using the Cox model without addressing the issue of missing values can lead to biased estimates or a lack of precision \citep{Chen:miscova:2018, Marshall2010, Hsu:non:para2019}.  

By definition, a missing value is an unobserved value that would be meaningful for the analysis if it were observed \citep{Little2020}. 
The literature distinguishes three main mechanisms of missing data: Missing Completely At Random (MCAR), Missing At Random (MAR), and Missing Not At Random (MNAR) \citep{molenberghs2014, Little2020, Tsiatis2006}.  

The MCAR assumption is often too restrictive to be realistic \citep{buuren2018}, while the MNAR assumption typically requires additional information about the reasons for the missing values \citep{molenberghs2014, Little2020}. In contrast, the MAR assumption relies only on observed data and is widely implemented in several statistical R packages such as \texttt{MICE} \citep{buuren2011}, \texttt{mi} \citep{gelman2011}, \texttt{MissForest} \citep{missforest:2012}, \texttt{jomo} \citep{Quartagno2023}.  

Under the MAR assumption, several families of methods allow for the estimation of Cox model parameters in the presence of missing covariate values. The first family relies on parametric methods. In this category, missing values are directly handled through a parametric model, such as the fully Bayesian approach or the maximum likelihood approach \citep{Chen:baye:miss2002,Chen:miscova:2018}. In the literature, the most flexible and accessible method is parametric Multiple Imputation (MI), which uses either sequential regression techniques \citep{buuren2011, buuren2018} or joint modeling \citep{carpenter2023,Quartagno2023}. The second family concerns semiparametric methods. In this method, complete observations are weighted to minimize potential bias caused by missing values \citep{molenberghs2014, Tsiatis2006}. Strictly positive weights are calculated from a parametric model of the missing data process \citep{Hsu:non:para2019, Seaman2013}. The \texttt{survival} \citep{Therneau2024} and \texttt{survey} \citep{lumley2024} packages in R allow for the implementation of IPW. The third family relates to nonparametric MI methods. Unlike parametric and semiparametric methods, the nonparametric method does not specify any parametric model for either the partially observed variable or the missing data process \citep{Hsu:non:para2019, molenberghs2014, Takeuchi:mi:2021}. It generally relies on specific criteria applied to the observed values to handle the missing data \citep{buuren2018, Little2020, Breiman:cart:1984}. Several \texttt{R} packages implement these approaches, such as decision trees with \texttt{missForest} \citep{missforest:2012} and nearest neighbor methods with \texttt{VIM} and \texttt{Hmisc} \citep{VIM:2016, hmisc:2024}.  

Under the assumption of non-informative right censoring for the time-to-event outcome, and assuming MAR mechanism, several authors have compared different families of methods within the Cox model framework.
\citet{Marshall2010,Ali2011} restricted their comparative study to variants of the parametric MI method. \citet{Yoo2018} highlighted that the performance of parametric MI methods can be influenced by the proportion of missing values and the type of partially observed variable. \citet{Seaman2013} pointed out that IPW is often preferable to parametric MI in contexts where multiple variables are missing for the same individual. The authors emphasize the overall efficiency of MI compared to IPW. \citet{Chen:nonpara:1999}  showed that using a nonparametric likelihood approach to handle missing categorical covariates in a Cox model leads to more efficient estimates than the approximate partial likelihood estimates and estimates from complete-case analysis, across various censoring scenarios.
\citet{Seaman:mi:iwp:2021}  propose an  approach in the context of linear regression that combines MI and IPW. MI is applied to selected covariates with isolated missing values, while IPW is used to adjust for the exclusion of observations with missing data in key variables that are not imputed. This strategy allows for maximal use of the available data while correcting for potential selection bias. Although originally motivated by different concerns, this logic echoes earlier ideas from \citet{Tsiatis2006}, who suggested that imputing targeted variables could help transform a nonmonotone missing data pattern into a monotone one, thus facilitating the use of IPW.
More recently, \citet{Thaweethai2021} and \citet{Bhattacharjee:miipw:2024} extended this line of work by integrating MI with IPW in frameworks aiming for robust or doubly robust inference. Notably, \citet{Little2024} explicitly refer to such combinations as “hybrid” approaches and show their potential advantages in heterogeneous missing data settings.


Our objective is to formalize a methodological framework for hybrid methods, considered as alternatives to conventional approaches for handling missing data in the Cox model. To our knowledge, few studies have structured these approaches within a unified framework that allows for assessing the impact of their combination on the bias and variance of estimators, depending on the degree of confidence placed in each component. This study aims to address this gap.

The paper is organized as follows. Section \ref{prelim:chap2} provides an overview of the Cox model and missing data mechanisms. Section \ref{classical:approach:chap2} presents the classical and popular methods for handling missing data under MAR. Section \ref{method:chap2} describes our proposed hybrid methods. Section \ref{simulation:chp2} introduces   a simulation study based on SUPPORT data (Study to Understand Prognoses and Preferences for Outcomes and Risks of Treatments) \citep{Knaus1995}.
Section \ref{application:chp2} illustrates an application of the different methods to this SUPPORT data. Finally, Section \ref{discussion:chap2} is dedicated to the discussion and main conclusions.

\section{Preliminairies} \label{prelim:chap2}
\subsection{Notation and Assumptions}

Let $(T, \delta, \mat{X})$ be a non-hierarchical survival data, where $T \geq 0$ represents the observed time (time to event or censoring), $\delta$ is an indicator variable defining the occurrence of the event ($\delta = 1$) or censoring ($\delta = 0$), and $\mat{X} = (X_1, X_2, \ldots, X_p)$ is a vector of $p$ independent variables, where $p$ denotes the total number of independent variables. For an individual $i$, the data are $(t_i, \delta_i, X_i)$, with $i = 1, 2, \ldots, n$, where $n$ represents the total number of observations.  We assume non-informative right censoring and that $X_1$ is a categorical variable with more than two category levels. This choice, although not restrictive for the proposed methods, is motivated by the application, which involves a partially observed categorical covariate. Except for $X_1$, all other variables are fully observed.  We denote by $\mat{X}_{\lbrace -1\rbrace} = (X_2, \ldots, X_p)$ the subvector of $\mat{X}$ excluding the partially observed covariate $X_1$. We assume that $X_1$ is missing according to a MAR mechanism.  Consider the partition $X_1 = {\lbrace X_1^{\obs}, X_1^{\mis} \rbrace}$, where $X_1^{\obs}$ and $X_1^{\mis}$ represent respectively the observed and missing parts of $X_1$. The missing data indicator variable $R$ is defined such that  $R = 1$ if $X_1 = X_1^{\obs}$ and $R = 0$ if $X_1 = X_1^{\mis}$ . Let $\dot{X}_1^{\mar} = {\lbrace X_1^{\obs}, X_1^{\mar} \rbrace}$, where $X_1^{\mar}$ denotes MI under MAR of $X_1^{\mis}$ by $f_{\{.\}}$, where $f_{\{.\}}$ is a parametric, nonparametric, or hybrid method. Let $U(\cdot)$ be the score function of the partial likelihood of the Cox model and $\beta$ the parameters of interest. Throughout the paper, we assume that the rules of \citet{Rubin1987} rules remain valid for any method involving MI.

\subsection{Review of the Cox Model}

The hazard function for an individual \(i\) with covariates \(\mat{X}_i\) is defined as:
\[
h(t \mid \mat{X}_i) = h_0(t) \exp(\beta^\top \mat{X}_i),
\]
where \(h_0(t)\) is an unspecified baseline hazard function, and \(\beta\) is a vector of regression coefficients \citep{Cox1972, Klein2012, Therneau:2000}.

In the absence of missing data, inference is typically based on the partial likelihood function, given by \citep{Kalbfl:prent:2011, Moore2016}:
\[
L(\beta) = \prod_{i=1}^{n} \left( \frac{\exp(\beta^\top \mat{X}_i)}{\sum_{j \in r(t_i)} \exp(\beta^\top \mat{X}_j)} \right)^{\delta_i},
\]
where \(r(t_i)\) denotes the risk set just before time \(t_i\), and \(\delta_i\) is the event indicator.

The corresponding score function is defined as:
\[
U(\beta) = \sum_{i=1}^{n} \delta_i \left( \mat{X}_i - \frac{\sum_{j \in r(t_i)} \mat{X}_j \exp(\beta^\top \mat{X}_j)}{\sum_{j \in r(t_i)} \exp(\beta^\top \mat{X}_j)} \right).
\]

The maximum partial likelihood estimator \(\hat{\beta}\) is obtained by solving 
$U(\hat{\beta}) = 0$.

\section{Review of classical missing data methods}\label{classical:approach:chap2}

This section discusses the commonly used methods for handling missing data within the Cox model framework under the assumption of non-informative right censoring. We do not aim to exhaustively review all existing methods but rather to explore popular techniques recognized for their ease of implementation and ability to provide reliable statistical inferences.

\subsection{Inverse Probability Weighting}

Under the assumption of non-informative right censoring, the m-estimator $\hat{\beta}_{\text{IPW}}$ is the solution to the score equation of the weighted partial likelihood \citep{Tsiatis2006}.

\[
 \sum_{i=1}^{n} \hat{w}_i  \left( \mat{X}_i - \frac{\sum_{j \in r(t_i)} \hat{w}_j \mat{X}_j \exp(\mat{X}_j^\top \hat{\beta}_{\text{IPW}})}{\sum_{j \in r(t_i)} \hat{w}_j \exp(\mat{X}_j^\top \hat{\beta}_{\text{IPW}})} \right)   = 0
\]

Where $\hat{w}_i = \frac{1}{\pi_i}$ with $\pi_i$ being the probability of observing individual $i$ given the fully observed covariates.
The variance  of $\hat{\beta}_{\text{IPW}}$ is obtained using the robust sandwich variance estimator \citep{Tsiatis2006, molenberghs2014}. The \texttt{coxph} function from the \texttt{survival} R package \citep{Therneau2024} already implements robust standard errors using the \texttt{weights} option. Additionally, the \texttt{sandwich} R package \citet{Zeileis:sandwich:2006}  provides a general implementation of the sandwich variance for various models.
In the IPW context, misspecification of the missing data process model can bias $\hat{\beta}_{\text{IPW}}$  \citep{Rubin1987,molenberghs2014}.

\subsection{Parametric MI method}

In the context of parametric MI, a consistent estimator $\hat{\beta}$ of the Cox model is obtained by solving \citep{carpenter2023}:

\begin{equation*}
E_{f_p(X_1^{\text{mis}} \mid T, X_1^{\text{obs}}, \mat{X}_{\lbrace -1\rbrace}, R)} \left\{ U\left(\hat{\beta}, T, \delta, X_1^{\text{obs}}, X_1^{\text{mis}}, \mat{X}_{\lbrace -1\rbrace} \right) \right\} = 0
\end{equation*}

Where $E_{f_p(.)}$ denotes the expectation of the score function of the partial likelihood conditional on a parametric imputation model \(f_p\). Parametric MI  techniques differ by the specification of   \( f_p(X_1^{\text{mis}} \mid T, X_1^{\text{obs}}, \mat{X}_{\lbrace -1\rbrace}, R) \). Two types of specifications are common in the literature: sequential regression and joint modeling. In sequential regression, each partially observed variable defines its own imputation model \citep{buuren2018, buuren2011}. In contrast, joint modeling considers all variables simultaneously in a joint distribution model (where \(f_p\) follows a multivariate normal distribution), allowing for the generation of consistent imputations by accounting for relationships between the variables \citep{carpenter2023, Quartagno2023}.

The general principle of MI  involves generating \( M \) complete data (\( m = 1, 2, 3, \dots, M \)) by replacing the missing values of \( X_1^{\mis} \) with plausible values \( X_1^{\mar} \) drawn from an appropriate imputation model. For example, under a parametric imputation model \( f_p(X_1^{\mis} \mid T, X_1^{\obs}, X_{\{-1\}}, R) \), \( M \) imputed datasets can be obtained. Then, each imputed data requires solving the following partial likelihood equation:
\[
U\left(\hat{\beta}_m, T, \dot{X}_1^{\text{imp}}, \mat{X}_{\lbrace -1\rbrace} \right) = \sum_{i=1}^{n} \delta_i \left( \mat{X}_i^{(m)} - \frac{\sum_{j \in r(t_i)} \mat{X}_j^{(m)} \exp(\hat{\beta}_m^\top \mat{X}_j^{(m)})}{\sum_{j \in r(t_i)} \exp(\hat{\beta}_m^\top \mat{X}_j^{(m)})} \right) = 0,\quad m = 1,\dots, M
\]
Where, \( \hat{\beta}_m \) denotes the estimate obtained from the \( m^{\text{th}} \) partial likelihood equation, and \( \dot{X} \) represents the set \( \{ \dot{X}_1^{\mar}, X_{\{-1\}} \} \). The \( M \) estimates are then combined using Rubin's rules \citep{Rubin1987}, yielding \( \hat{\beta}_{\text{MI}} \) as the pooled estimate and \( \text{Var}(\hat{\beta}_{\text{MI}}) \) as the corresponding variance estimate for MI.
\[
\hat{\beta}_{\text{MI}} = \frac{1}{M} \sum_{m=1}^{M} \hat{\beta}_m, \quad 
\text{Var}(\hat{\beta}_{\text{MI}}) = \frac{1}{M} \sum_{m=1}^{M} \widehat{\text{Var}}(\hat{\beta}_m) 
+ \left(1 + \frac{1}{M} \right) \cdot \left[ \frac{1}{M-1} \sum_{m=1}^{M} (\hat{\beta}_m - \hat{\beta}_{\text{MI}})^2 \right]
\]
Misspecification of the parametric imputation model \( f_p(X_1^{\text{mis}} \mid T, X_1^{\text{obs}}, X_{\{-1\}}, R) \) can lead to biased estimates of \(\hat{\beta}_{\text{MI}}\) \citep{Rubin1987}. Moreover, parametric MI has several practical limitations. First, it cannot accommodate more predictor variables than observations without relying on informative prior distributions. Second, the inclusion of highly correlated predictors may lead to instability in the imputation model due to multicollinearity  \citep{Shah2014}.

\subsection{Nonparametric MI method}

In the context of nonparametric MI, a convergent estimator $\hat{\beta}$ of the Cox model is obtained by solving \citep{carpenter2023}:

\begin{equation*}
E_{f_{np}(X_1^{mis} | T, X_1^{obs},X_{ \{ -1 \} }, R)} \left\{ U\left(\hat{\beta}, T, \delta, X_1^{obs},X_1^{mis},X_{ \{ -1 \} } \right) \right\} = 0
\end{equation*}

Where $E_{f_{np}(.)}$ denotes the expectation of the partial likelihood score function conditionally on a nonparametric imputation model \( f_{np} \).

Nonparametric multiple imputation methods relax the need for strong parametric assumptions when estimating the conditional distribution \( f_{np}(X_1^{\text{mis}} \mid T, X_1^{\text{obs}}, X_{\{-1\}}, R) \). While potentially less efficient than parametric approaches particularly in small samples, they offer enhanced robustness and reduced bias in settings where the underlying relationships are nonlinear, involve complex interactions, or violate classical distributional assumptions. Instead, they typically use specific criteria applied to the observed values to handle missing data \citep{buuren2018, Breiman:cart:1984}.

Several R packages support nonparametric imputations, including Classification And Regression Trees (CART) \citep{buuren2018, Breiman:cart:1984}, Decision trees with \texttt{missForest} \citep{missforest:2012}, Nearest neighbor methods with \texttt{VIM} and \texttt{Hmisc} \citep{VIM:2016, hmisc:2024}. Additionally, other nonparametric imputation methods, such as Hot-deck and cold deck methods, are discussed in the literature. For an in-depth review of the imputation principles underlying each method, see \citep{Rubin1987, molenberghs2014,buuren2018,Breiman:cart:1984}. Despite their flexibility, nonparametric methods also present several limitations. Both CART and random forests are conservative in nature, often shrinking estimates toward zero, and they may struggle to accurately capture linear main effects \citep{Shah2014}. These methods can yield greater variability across imputations, particularly in small datasets. Moreover, they typically lack an explicit probabilistic model for uncertainty, which complicates theoretical justification. Rubin's rule applies in the same way as in the previous section.

\section{Hybrid methods}\label{method:chap2}

In this section, we propose hybrid methods whose primary objective is to leverage the strengths of multiple classical approaches in order to mitigate the limitations inherent to each method when used in isolation for handling missing data. This section also provides the theoretical and practical justifications for the proposed combinations.

\subsection{Hybrid methods procedure}

Hybrid methods integrate at least two classical methods of handling missing data into a unified procedure. We illustrate a four hybrid methods and their implementation in the context of the semiparametric Cox model.

\subsubsection*{Hybrid~1: Combination of Parametric and Nonparametric MI}

This hybrid approach combines multiple imputations generated under the MAR assumption using two different models: a parametric model \( f_p(X_1^{\text{mis}} \mid T, X_1^{\text{obs}}, X_{\{-1\}}, R) \), and a nonparametric model \( f_{np}(X_1^{\text{mis}} \mid T, X_1^{\text{obs}}, X_{\{-1\}}, R) \). Let \( M_1 \) denote the number of imputations drawn from the parametric model and \( M_2 \) from the nonparametric model, such that the total number of imputations is \( M = M_1 + M_2 \). The values of \( M_1 \) and \( M_2 \) can be adapted to reflect the desired balance between model-based efficiency and robustness.
 The Cox model is then analyzed on the entire set of $ M$ imputed data, and the results are combined according to the rule of \citep{Rubin1987}. This hybrid method stands out for its flexibility, exploiting the advantages of both parametric and nonparametric imputation techniques to better capture the plausibility of missing data. 

\subsubsection*{Hybrid~2: Parametric MI and Semiparametric (IPW).}

Hybrid~2 combines, under the MAR assumption, a parametric imputation model\\
\( f_p(X_1^{\mis} \mid T, X_1^{\obs}, \mat{X}_{\{-1\}}, R) \) to generate \( M \) imputed datasets, and a model for the missing data mechanism to compute individual weights. Unlike standard IPW weights, we propose an adjusted weighting scheme that accounts for both the contribution of missing observations and the sampling variability. In other words, for an individual \( i \), we define these weights as:

\[
\hat{w}_i =
\begin{cases} 
\frac{1}{\pi_i}, & \text{if } R_i = 1, \\
\kappa \cdot 1 + (1-\kappa) \cdot \left( \frac{1}{1 - \pi_i} \right), & \text{if } R_i = 0,
\end{cases}
\]

where \(\kappa \in [0,1]\) is a compromise parameter between the weight associated with each observation (which is 1) in the imputed data and the weight associated with the missing observations (which is \( \frac{1}{1 - \pi_i} \)). Specifically, the compromise parameter,  represents the weight assigned to the MI approach relative to IPW. Its variation allows the assessment of the impact of this weighting on the bias and variance of the estimators, with the aim of identifying an optimal value that achieves a satisfactory trade-off.


The Cox model, which serves as our final analysis model, is then fitted on each of the \(M\) imputed datasets, incorporating the previously defined observation weights. For each \(m = 1, 2, 3, \dots, M\), the variance of the estimated coefficients is computed using robust sandwich-type estimators, as recommended when weights are included in the estimation process \citep{Tsiatis2006, Seaman2013}
Finally, Rubin’s rule is applied to combine the estimates and calculate their variances \citep{Rubin1987}. 

The Cox model, which serves as our final analysis model, is then fitted to each of the \(M\) imputed datasets, incorporating the previously defined observation weights. For each \(m = 1, 2, 3, \dots, M\), the variance of the estimated coefficients is computed using robust sandwich-type estimators, as recommended when weights are included in the estimation process \citep{Tsiatis2006}. Finally, Rubin’s rule is applied to combine the estimates and compute the overall variance \citep{Rubin1987}. Unlike Hybrid~2, Hybrid~1 does not rely on specifying a compromise parameter to balance the contribution of its components.

\subsubsection*{Hybrid~3: Nonparametric MI and IPW}

Hybrid 3  follows the same principle as Hybrid 2. The main difference lies in the use of nonparametric MI, \( f_{np}(X_1^{\mis} | T, X_1^{\obs}, X_{\{-1\}}, R) \), in place of parametric MI, \( f_p(X_1^{\mis} | T, X_1^{\obs}, \mat{X}_{\{-1\}}, R) \).  
The  use of weights remain identical to those described in Hybrid~2. Rubin’s rule is still applied to obtain the estimates and variances.

\subsubsection*{Hybrid~4: Parametric MI, Nonparametric MI, and IPW}

Hybrid~4 method combines, under the MAR assumption, \(M/2\) imputed data from a parametric MI model, \( f_p(X_1^{\mis} | T, X_1^{\obs}, X_{\{-1\}}, R) \) and \(M/2\) imputed data from a nonparametric MI model, \( f_{np}(X_1^{\mis} | T, X_1^{\obs}, \mat{X}_{\{-1\}}, R) \). The entire \(M/2 + M/2 = M\) imputed data are analyzed using the Cox model, accounting for the weights of the observations defined in Hybrid~2, and the results are then combined according to Rubin’s rule.

\subsection{Hybrid methods justifications}

The motivation for introducing hybrid methods under MAr is based on several important considerations.

 \textbf{i. Absence of a single ideal method.} There is no single, ideal imputation method to replace missing values with plausible values \citep{carpenter2023, Little2020}.

 \textbf{ii. Complexity of real data.} It is often prudent to combine several potential methods to better capture the uncertainties of missing values in real data as suggested by combining MI and IPW approaches \citep{Seaman:mi:iwp:2021}. This strategy aligns with the recommendations of \citet{Rubin1987}, who emphasizes the importance of modeling the variability inherent in the imputation of missing data.

 \textbf{iii. Reduction of model-related weaknesses.}  
Hybrid methods help mitigate the limitations of both parametric and nonparametric multiple imputation. They provide a flexible framework that balances bias and variance trade-offs across classical approaches. On one hand, they reduce the impact of misspecification in parametric imputation models. On the other hand, they help counter the high variability that can affect nonparametric methods, especially in small samples. For example, the nonparametric MI component captures nonlinear relationships, interactions, multicollinearity, and influential or outlier values \citep{Sela2012,Breiman:cart:1984}.

 \textbf{iv. Independence of Rubin's rule.}
Rubin’s rule, used to combine results, is independent of the choice of the imputation model and the final analysis \citep{carpenter2023,buuren2018}. This provides flexibility in the choice and combination of classical methods.

\section{Simulation study} \label{simulation:chp2}

This simulation study is based on real data from the SUPPORT study \citep{Knaus1995}, available in the \texttt{Hmisc} package (version 5.2.3) \citep{hmisc:2024} . The original dataset includes \(n = 9105\) observations, with missing values on several explanatory covariates, particularly the \texttt{income} variable. A complete-case subsample of \(6083\) individuals was extracted and used as the reference dataset. The retained variables include survival time, death indicator, sex, income, respiratory rate (\texttt{resp}), serum sodium and level (\texttt{sod}). This selection, driven by methodological rather than clinical considerations, aimed to retain variables likely to meet the assumptions of the Cox model while preserving valid clinical interpretability. Moreover, using a large reference population ensures the reliable estimation of performance criteria in our realistic simulation framework. The proportion of censored observations is 33\%, with death considered as the event and being alive at the end of follow-up considered as censoring.

From this reference dataset, \(1000\) independent subsamples of size \(n = 1000\) were drawn without replacement, in order to preserve the representativeness of covariates, especially categorical variables. Unlike approaches using sampling with replacement \citep{Shah2014}, our strategy reduces the correlation among simulated observations and avoids imbalance in category representation, thereby ensuring a more realistic simulation that mimics empirical data.

The assumptions of linearity and proportional hazards were checked: age showed a marked non-linear effect, and both \texttt{sex} and \texttt{age} exhibited slight violations of the proportional hazards assumption. A  MAR mechanism was artificially introduced on the \texttt{income} variable using the \texttt{ampute()} function from the \texttt{mice} package \citep{buuren2018}, based on a missingness pattern conditional on \texttt{age}, \texttt{death}, and \texttt{sod}, with a missingness rate of 30\%. 

The number of imputations was set to \(M = 10\) to balance computational efficiency and estimation stability. The imputation models primarily followed a parametric framework, using regression models tailored to each variable type—logistic regression for binary variables, Bayesian linear regression for continuous variables, and multinomial regression for categorical variables. The cumulative hazard function was estimated using the Nelson–Aalen estimator, and event status was included as a covariate in both parametric and nonparametric imputation models, as recommended by \citet{White:cox:2009}. 

In addition, a nonparametric approach based on random forests was employed, given its demonstrated robustness in handling complex relationships and interactions \citep{Sela2012}. Both classical and hybrid methods were compared. Performance was evaluated using absolute and relative bias, the average width of the $95\%$ confidence intervals, and nominal coverage rates. Several values of the compromise parameter \(\kappa\) were examined to assess the behavior of the weighted hybrid methods. Estimates derived from the complete dataset were used as the reference benchmark for comparative evaluation. All simulations and analyses were conducted using \texttt{R} version 4.5.0. We used the \texttt{survival} package (version 3.8-3) for the final analyses. For multiple imputation methods, both parametric and non-parametric, we used the \texttt{mice} package (version 3.17.0).

\section{Simulation results} \label{simu:reslt:chp2}

\begin{figure}[H]  
    \centering
    \includegraphics[width=0.85\textwidth]{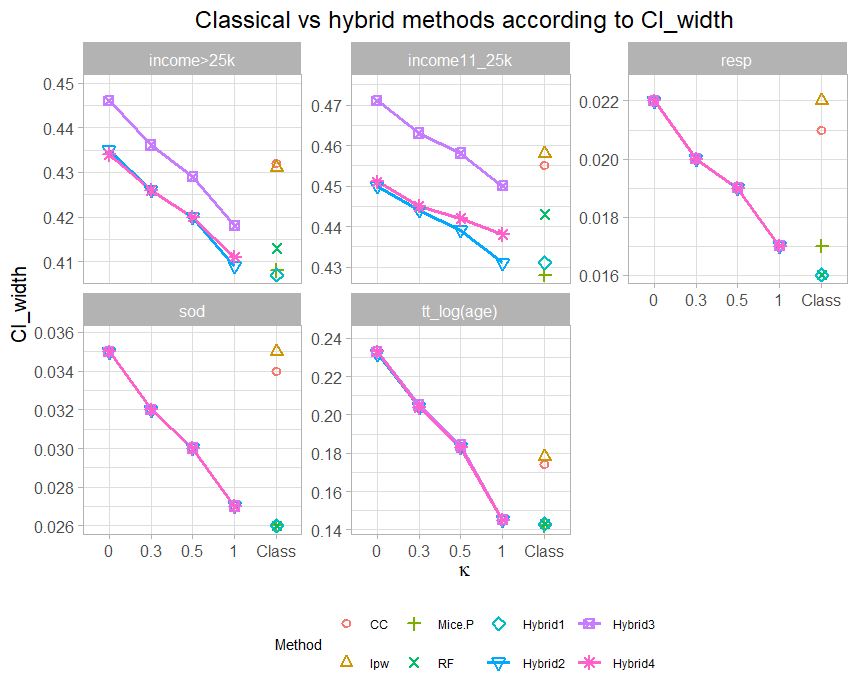}
  \caption{Relative bias (\%) and average 95\% confidence interval width of classical and hybrid methods as a function of \(\kappa\) (applicable only to hybrid methods), based on simulated data from the SUPPORT study.}
   \label{support_biais_CI:chp2}
  \end{figure}

\begin{figure}[H]  
    \centering
    \includegraphics[width=0.85\textwidth]{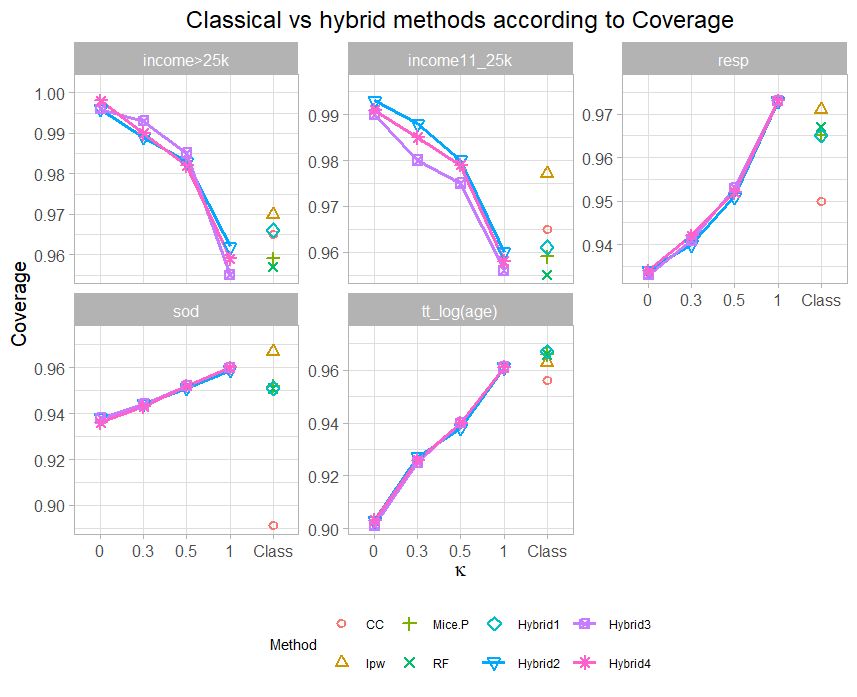}
  \caption{Empirical 95\% coverage rates of classical and hybrid methods as a function of \(\kappa\) (relevant only for hybrid methods), based on simulated data from the SUPPORT study.}
   \label{support_coverage:chp2}
  \end{figure}

Figures~\ref{support_biais_CI:chp2} and~\ref{support_coverage:chp2} display, respectively, the relative bias, the average width of the 95\% confidence intervals, and the empirical coverage rates for both classical and hybrid methods, as a function of the compromise parameter \(\kappa\).

\textbf{i. Fully observed variables.}  
For variables without missing values (respiratory rate, sodium, and log-transformed age), relative biases are generally close to zero across all methods, with the notable exception of the complete-case approach, which exhibits a severe bias for sodium (+166\%). The narrowest confidence intervals are obtained with MICE-P, RF, and Hybrid~1, while coverage remains satisfactory for most methods—except for complete cases and very low values of \(\kappa\). This confirms that classical methods are generally reliable when applied to fully observed variables.

\textbf{ii. Partially observed variable: income.}  
In contrast, differences across methods are much more pronounced for this partially observed variable. The complete-case approach results in substantial biases (up to -94.4\%), low precision, and poor coverage. Although MICE-P reduces uncertainty, it remains heavily biased (ranging from -84\% to -14\%), while RF tends to overestimate effects (+30\% to +35\%). Only IPW achieves moderate bias (7.4\% and 15.9\%), but with a lower coverage compared to the best-performing hybrid methods.

Hybrid approaches 2 to 4 outperform classical methods across all evaluation criteria, particularly for \(\kappa \in [0.3, 0.5]\). At \(\kappa = 0.5\), Hybrid~3 is the only method achieving a bias below 5\% for both income categories (-3.5\% and +3.1\%), with moderate interval widths and coverage rates exceeding 98\%. Hybrid~4 also demonstrates systematically lower bias than any classical method (e.g., -12.6\% and +1.5\%), along with high coverage and balanced precision.

At the extremes of \(\kappa\), performance deteriorates: for \(\kappa = 1\), biases in Hybrid~3 can exceed +45\%, whereas for \(\kappa = 0\), biases become strongly negative and coverage rates drop, particularly for the income (>25k) category. These trends underscore the importance of a well-calibrated compromise parameter.

\textbf{iii. Robustness of Hybrid~4.}  
Hybrid~4, which combines MICE-P, RF, and IPW, exhibits remarkable stability across variables and \(\kappa\) values. Its intermediate position between Hybrid~1 (MICE-P + IPW) and Hybrid~2 (RF + IPW) grants it robustness against potential misspecification of the parametric and non-parametric imputation models, as well as the missingness model. It consistently maintains coverage above 95\% while keeping the bias lower than any individual method alone. For further details, see Tables~\ref{tab:coverage_support}, \ref{tab:bias_support}, and~\ref{tab:ciwidth_support}.

\begin{figure}[H]  
    \centering
    \includegraphics[width=0.85\textwidth]{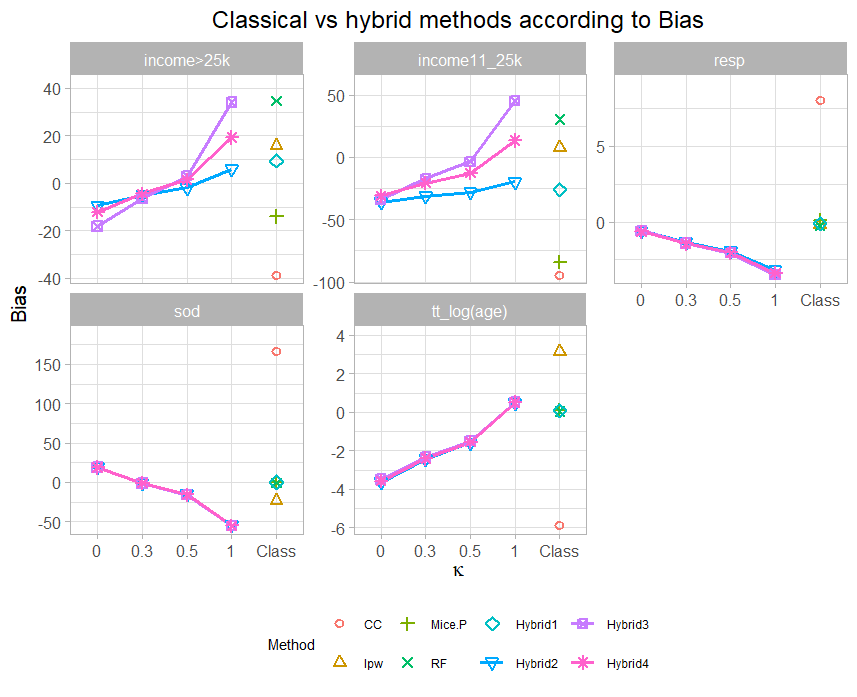}
  \end{figure}

\section{Application}\label{application:chp2}

As part of the proposed hybrid methods, we use data from the SUPPORT study (Study to Understand Prognoses and Preferences for Outcomes and Risks of Treatments) \citep{Knaus1995}, a prospective cohort conducted in five medical centers in the United States. The study was carried out in two phases: the first (1989–1991) included 4 301 patients, and the second (1992–1994) included 4 804 patients. Eligible patients were hospitalized for severe conditions associated with a poor short-term prognosis. Patients who died shortly after admission or for whom no usable follow-up data were available were excluded from the analyses. No significant differences were observed in the characteristics of patients between the two phases \citep{Knaus1995}.  The data are available through the \texttt{Hmisc} package \citep{hmisc:2024} and have been used by several authors \citep{Harrell2015,Ren2019,Zhang2022}. 
Our objective is to evaluate the association between patients' income and the risk of death.

\subsection*{SUPPORT Data}  

The analysis is restricted to individuals with missing values for the exposure variable of interest, \texttt{income}, while all other variables—such as survival time, death indicator, sex, respiratory rate (\texttt{resp}), and serum sodium level (\texttt{sod}) are fully observed. Notably, the same set of variables was used in our realistic simulation framework to ensure consistency in the analytical structure, although the simulated datasets are based on repeated samples drawn from the complete-case subset (see Table~\ref{tab:characteristics} for the characteristics of the real data).
 
The final sample consists of 9 039 patients ($99.27\%$ of patients from phases 1 and 2), of whom $31.97\%$ are censored. The proportion of missing values for income is $32.70\%$. Given the absence of substantive evidence suggesting an alternative missing data mechanism for the variable \texttt{income}, we adopt the MAR assumption. This choice aligns with recommendations from the literature, where MAR is often considered a reasonable starting point for missing data analysis \citep{Rubin1987, carpenter2023}.

In applying the proposed hybrid methods, we verified the key assumptions of the Cox model, namely linearity and proportional hazards. For continuous covariates, several functional forms were evaluated to assess linearity, and the most appropriate transformations were applied where necessary. The final specifications were selected based on a combination of statistical fit and parsimony, and are consistent with the assumptions used in the simulation framework. Regarding the proportional hazards assumption, interactions with time were introduced for continuous covariates when violations were detected, and stratified Cox models were used for categorical variables as appropriate. Further details on these assessments and modeling decisions are provided in Appendix~\ref{linearity:PH}.

\begin{table}[H]
\centering

\caption{Patient characteristics stratified by death status in the subsample of the SUPPORT data}
\label{tab:characteristics}
{
\setlength{\tabcolsep}{16pt}
\begin{tabular}{lcc}
\toprule
\rowcolor{gray!40} \textbf{Variable} & \textbf{Survivors (n = 2890)} & \textbf{Deaths (n = 6149)} \\
\midrule
\textbf{Follow-up (quarters, mean (SD))} & 11.62 (5.65) & 2.27 (3.53) \\
\rowcolor{gray!40} \textbf{Age (mean (SD))} & 58.57 (16.68) & 64.58 (14.69) \\
 \textbf{Sex (Male, n~(\%))} &  1561 (54.0) & 3526 (57.3) \\
\rowcolor{gray!40} \textbf{Income, n (\%)} & & \\
\hspace{1cm} <$11k$     & 932 (45.9) & 1906 (47.0) \\
\hspace{1cm} $11$–$25k$ & 503 (24.8) & 1007 (24.8) \\
\hspace{1cm} >$25k$     & 594 (29.3) & 1141 (28.1) \\
\rowcolor{gray!40}  \textbf{Respiratory rate (mean (SD))} & 23.35 (9.48) & 23.57 (9.36) \\
 \textbf{Sodium (mean (SD))}           & 137.73 (5.87) & 137.47 (6.04) \\
\bottomrule
\end{tabular}
}
\end{table}

\begin{table}[H]
\centering
\caption{Patients characteristics stratified by missingness indicator (R) for income in the subsample of the SUPPORT data} \label{tab:R_income}
{
\setlength{\tabcolsep}{16pt}
\begin{tabular}{lcc}
\toprule
\rowcolor{gray!40} \textbf{Variable} & \textbf{R = 0 (n = 2956)} & \textbf{R = 1 (n = 6083)} \\
\midrule
\textbf{Death (Yes, n~(\%))}  &  2095 (70.9) & 4054 (66.6) \\
\rowcolor{gray!40} \textbf{Sex (Male, n~(\%))} &  1671 (56.5) & 3416 (56.2) \\
\textbf{Follow-up time, quarters (mean (SD))} & 5.02 (6.22) & 5.37 (6.10) \\
\rowcolor{gray!40} \textbf{Age       (mean (SD))}                 & 64.05 (15.70) & 61.98 (15.52) \\
\textbf{Sodium     (mean (SD))}                & 137.72 (5.91) & 137.47 (6.03) \\
\rowcolor{gray!40} \textbf{Respiratory rate   (mean (SD))}        & 23.62 (9.67)  & 23.44 (9.27) \\
\bottomrule
\end{tabular}
}
\end{table}

\subsection*{Results and Interpretation} 

In line with the conclusions of the simulation presented in Section~\ref{simu:reslt:chp2}, Hybrid~3 (RF + IPW) emerged as the most balanced approach in terms of performance  for values of \(\kappa \in [0.3, 0.5]\), whereas Hybrid~4 (MICE + RF + IPW) exhibited greater overall stability. Consequently, the interpretation of the results from the real data (see Section~\ref{result:support_data_subset}) primarily relies on these two methods. For the intermediate income category (\$11k–\$25k), the estimated hazard ratios (HRs) range from 0.9844 (HYBRID~3, \(\kappa = 0.5\)) to 0.9863 (HYBRID~4, \(\kappa = 0.3\)), with confidence intervals (CIs) ranging from [0.9109; 1.0655] to [0.9156; 1.0614]. For the higher income category (\(>\$25k\)), HRs lie between 0.9777 and 0.9798, with CIs between [0.9103; 1.0532]. All intervals include 1, indicating no statistically significant association between higher income and mortality risk. Classical methods—namely IPW, parametric MICE, RF, and complete case analysis—yield very similar results. For the \$11k–\$25k category, HRs range from 0.9820 (IPW) to 0.9852 (MICE), with CIs from [0.9092; 1.0606] to [0.9193; 1.0558]. For the \(>\$25k\) category, HRs range from 0.9754 (IPW) to 0.9802 (MICE), with CIs from [0.9058; 1.0503] to [0.9134; 1.0518]. Hybrid~1 (MICE + RF) and Hybrid~2 (MICE + IPW) also provide comparable estimates, regardless of \(\kappa\), with HRs around 0.985 and confidence intervals that consistently include 1. Overall, the results across all methods converge on the same conclusion: there is no statistically significant association between income level and mortality risk in this subsample of the SUPPORT data.

{
\renewcommand{\arraystretch}{0.8} 
\begin{center}
{
\setlength{\tabcolsep}{16pt}
\begin{longtable}{lccccc}
\caption{Hazard ratios (HR) with 95\% confidence intervals (lower and upper bounds) for classical and hybrid methods, for selected covariates in the SUPPORT dataset.}
\label{result:support_data_subset}\\

\rowcolor{gray!40} \multirow{2}{*}{\textbf{Methods}} & \multicolumn{2}{c}{\textbf{Income}} & \textbf{Sodium} & \textbf{Respiration} & \textbf{tt(logage)} \\
\cmidrule(r){2-3} \cmidrule(r){4-4} \cmidrule(r){5-5} \cmidrule(r){6-6}
 & $11$k–$25$k & $>$25k & & & \\
\hline
\endfirsthead

\rowcolor{gray!40} \multirow{2}{*}{\textbf{Methods}} & \multicolumn{2}{c}{\textbf{Income}} & \textbf{Sodium} & \textbf{Respiration} & \textbf{tt(logage)} \\
\cmidrule(r){2-3} \cmidrule(r){4-4} \cmidrule(r){5-5} \cmidrule(r){6-6}
 & $11$k–$25$k & $>$25k & & & \\
\hline
\endhead

\hline
\multicolumn{6}{c}{Continued on next page}\\
\hline
\endfoot

\hline
\endlastfoot
\rowcolor{gray!40} \textbf{CC} & & & & & \\
\hspace{0.2cm} HR         & 0.9830 & 0.9760 & 0.9960 & 1.0029 & 1.2205 \\
\hspace{0.2cm} CI lower   & 0.9103 & 0.9064 & 0.9907 & 0.9995 & 1.1820 \\
\hspace{0.2cm} CI upper   & 1.0615 & 1.0509 & 1.0014 & 1.0064 & 1.2602 \\

\rowcolor{gray!40} \textbf{IPW} & & & & & \\
\hspace{0.2cm} HR         & 0.9820 & 0.9754 & 0.9965 & 1.0029 & 1.2177 \\
\hspace{0.2cm} CI lower   & 0.9092 & 0.9058 & 0.9909 & 0.9993 & 1.1799 \\
\hspace{0.2cm} CI upper   & 1.0606 & 1.0503 & 1.0022 & 1.0064 & 1.2566 \\

\rowcolor{gray!40} \textbf{MICE (P)} & & & & & \\
\hspace{0.2cm} HR         & 0.9852 & 0.9802 & 0.9969 & 1.0035 & 1.2007 \\
\hspace{0.2cm} CI lower   & 0.9193 & 0.9134 & 0.9926 & 1.0007 & 1.1700 \\
\hspace{0.2cm} CI upper   & 1.0558 & 1.0518 & 1.0013 & 1.0063 & 1.2322 \\

\rowcolor{gray!40} \textbf{RF} & & & & & \\
\hspace{0.2cm} HR         & 0.9823 & 0.9758 & 0.9969 & 1.0035 & 1.2006 \\
\hspace{0.2cm} CI lower   & 0.9151 & 0.9135 & 0.9926 & 1.0007 & 1.1699 \\
\hspace{0.2cm} CI upper   & 1.0545 & 1.0422 & 1.0013 & 1.0063 & 1.2320 \\
\hline
\multicolumn{6}{c}{\textbf{Independent of $\kappa$}} \\
\rowcolor{gray!40} \textbf{HYBRID 1} & & & & & \\
\hspace{0.2cm} HR         & 0.9852 & 0.9789 & 0.9969 & 1.0035 & 1.2006 \\
\hspace{0.2cm} CI lower   & 0.9179 & 0.9160 & 0.9926 & 1.0007 & 1.1699 \\
\hspace{0.2cm} CI upper   & 1.0574 & 1.0463 & 1.0013 & 1.0063 & 1.2321 \\

\hline
\multicolumn{6}{c}{\textbf{For $\kappa = 0.3$}} \\

\rowcolor{gray!40} \textbf{HYBRID 2} & & & & & \\
\hspace{0.2cm} HR         & 0.9852 & 0.9799 & 0.9970 & 1.0036 & 1.1955 \\
\hspace{0.2cm} CI lower   & 0.9130 & 0.9042 & 0.9923 & 1.0007 & 1.1628 \\
\hspace{0.2cm} CI upper   & 1.0632 & 1.0620 & 1.0017 & 1.0066 & 1.2292 \\

\rowcolor{gray!40} \textbf{HYBRID 3} & & & & & \\
\hspace{0.2cm} HR         & 0.9852 & 0.9783 & 0.9970 & 1.0036 & 1.1955 \\
\hspace{0.2cm} CI lower   & 0.9109 & 0.9103 & 0.9923 & 1.0007 & 1.1629 \\
\hspace{0.2cm} CI upper   & 1.0655 & 1.0513 & 1.0017 & 1.0066 & 1.2291 \\

\rowcolor{gray!40} \textbf{HYBRID 4} & & & & & \\
\hspace{0.2cm} HR         & 0.9863 & 0.9798 & 0.9970 & 1.0036 & 1.1955 \\
\hspace{0.2cm} CI lower   & 0.9132 & 0.9114 & 0.9923 & 1.0007 & 1.1628 \\
\hspace{0.2cm} CI upper   & 1.0653 & 1.0532 & 1.0017 & 1.0066 & 1.2291 \\

\hline
 \multicolumn{6}{c}{\textbf{For $\kappa = 0.5$}} \\
\rowcolor{gray!40} \textbf{Hybrid2} & & & & & \\
\hspace{0.2cm} HR         & 0.9849 & 0.9797 & 0.9970 & 1.0035 & 1.1972 \\
\hspace{0.2cm} CI lower   & 0.9156 & 0.9078 & 0.9924 & 1.0006 & 1.1653 \\
\hspace{0.2cm} CI upper   & 1.0595 & 1.0573 & 1.0017 & 1.0065 & 1.2299 \\

\rowcolor{gray!40} \textbf{Hybrid3} & & & & & \\
\hspace{0.2cm} HR         & 0.9844 & 0.9777 & 0.9970 & 1.0035 & 1.1971 \\
\hspace{0.2cm} CI lower   & 0.9133 & 0.9123 & 0.9924 & 1.0006 & 1.1653 \\
\hspace{0.2cm} CI upper   & 1.0611 & 1.0477 & 1.0017 & 1.0065 & 1.2298 \\

\rowcolor{gray!40} \textbf{Hybrid4} & & & & & \\
\hspace{0.2cm} HR         & 0.9858 & 0.9794 & 0.9970 & 1.0035 & 1.1971 \\
\hspace{0.2cm} CI lower   & 0.9156 & 0.9136 & 0.9924 & 1.0006 & 1.1653 \\
\hspace{0.2cm} CI upper   & 1.0614 & 1.0499 & 1.0017 & 1.0065 & 1.2298 \\
\hline
\end{longtable}
}
\end{center}
}
\vspace{-1cm}
\begin{tablenotes}
\scriptsize
\item  \textbf{CC}: complete case analysis; \textbf{IPW}: inverse probability weighting; \item \textbf{MICE (P)}: parametric imputation using the \texttt{polyreg} method; \textbf{RF}: non parametric imputation using random forests.  
\item \textbf{HYBRID 1}: combination of MICE (P) and RF; \textbf{HYBRID 2}: combination of MICE (P) and IPW.
\item \textbf{HYBRID 3}: combination of RF and IPW; \textbf{HYBRID 4}: combination of MICE (P), RF, and IPW.
\end{tablenotes}

\section{Discussion}\label{discussion:chap2}  
\textbf{Complexity of real data in the presence of missing values in analysis.} 
When handling missing data under the MAR assumption, many studies compare classical methods within the Cox model framework \citep{Marshall2010, Ali2011, Yoo2018, Seaman2013}. However, several questions remain: Is parametric MI an ideal benchmark? Is nonparametric MI  adequate to effectively handle missing values? The same question applies to semiparametric methods. The answer is not always straightforward. As highlighted by \citet{Little2020}, just as a single imputed value may fail to capture the full spectrum of plausible values, relying on a single classical method may also fall short in addressing the complexity of real data, particularly when the missingness mechanism is uncertain or the model is misspecified. To overcome these limitations, hybrid methods that combine the strengths of existing approaches are emerging as a compelling alternative \citep{Bhattacharjee:miipw:2024, Seaman:mi:iwp:2021, Little2024}. While \citet{Tsiatis2006} and \citet{Seaman:mi:iwp:2021} do not propose a hybrid formulation involving a tuning parameter \(\kappa\), they do underscore the complementarity between MI and IPW, which has inspired the development of combined strategies. Our study contributes to this evolving literature by systematically evaluating such hybrid approaches, focusing on the role and impact of \(\kappa\), using both simulation experiments and real data analysis.

\textbf{Comparison with Previous Work.}  
Our findings align with prior work pointing out the limitations of relying solely on one method. Parametric MI is efficient when the imputation model is correctly specified, but its performance deteriorates under misspecification \citep{Marshall2010, buuren2018}. Nonparametric methods, such as random forest imputation, are more flexible \citep{Yoo2018}, but they tend to produce higher variance \citep{Shah2014}. IPW, for its part, is highly sensitive to the specification of the missingness model and may yield unstable estimates, particularly in small samples or under high missingness \citep{Tsiatis2006, Seaman2013}. The hybrid framework we propose complements this body of work by offering a structured and interpretable mechanism to blend MI and IPW through the parameter \(\kappa\), which reflects the relative confidence assigned to each method.

\textbf{Strengths of the Study.}  
One major strength of our study lies in the generality and adaptability of the hybrid framework, which enables transparent sensitivity analyses across various methodological configurations. The use of realistic data-generating processes in our simulations, alongside the application to the SUPPORT dataset, reinforces the practical relevance of the findings. Furthermore, hybrid approaches offer strong flexibility and robustness by combining methods rooted in different statistical paradigms. They are also accessible in practice, thanks to R packages such as \texttt{mice}, \texttt{jomo}, \texttt{mi}, \texttt{randomForest}, and \texttt{missForest} \citep{buuren2011, Quartagno2023, gelman2011, missforest:2012}. In our study, this flexibility was embodied in four hybrid strategies (Hybrid~1 to Hybrid~4), each designed to accommodate specific data structures and analytic goals. All R code is available on our GitHub page \url{https://github.com/abdoulaye-dioni/Hybrid_methods}

\textbf{Limitations of the Study.}  
Despite its contributions, this study has several limitations. First, our simulations were conducted under a MAR assumption that we explicitly controlled, but in the real data analysis, we assumed MAR without being able to verify it. As widely noted in the literature, MAR is inherently untestable using the observed data alone \citep{molenberghs2014, carpenter2023}.

Additionally, we did not observe statistically significant differences between classical and hybrid approaches in the SUPPORT application. Several factors may explain this. To ensure methodological clarity, we employed a relatively simple analysis model in the simulations—potentially excluding covariates that strongly predicted missingness. In cases where nonlinear transformations offered only marginal gains over linear ones, we opted for linear models to prioritize our evaluation of hybrid methodology. Moreover, it is plausible that the missingness mechanism in the SUPPORT subsample approximates MCAR, which would limit the benefits of MAR-based methods (both classical and hybrid), except for a potential gain in precision.

Another important limitation concerns the computational burden: hybrid methods involving IPW are substantially more demanding than standard approaches or hybrid methods that rely solely on MI-based strategies. Moreover, probabilities close to 0 or 1 were truncated to lie within the \([0.01, 0.99]\) range to stabilize the weights, as such extreme values can induce high variability in the weights, leading to unstable estimates \citep{White2010,Tsiatis2006}.

We also assumed the validity of Rubin’s rules for variance combination without providing a formal proof in the hybrid context. While the imputation methods employed are theoretically proper in Rubin’s sense \citep{Rubin1987, buuren2018}, the hybrid framework may introduce additional complexities that warrant further justification. Moreover, we did not examine in detail the potential impact of misspecifying the imputation model in the semiparametric Cox setting, which could influence the accuracy of parameter estimation \citep{White:cox:2009}.

Finally, our simulation design allowed individuals to appear in multiple replicates. While this design choice preserved the empirical structure of the data, it introduced inter-replicate dependencies that we treated as negligible. This simplification may have led to a slight underestimation of the true variability across replications.

\textbf{Perspectives for Future Research.}  
Several extensions can be envisioned. First, it would be valuable to formally derive Rubin’s variance combination rule in the hybrid setting to reinforce its theoretical validity. Second, although Hybrid~4 consistently outperformed other strategies empirically, we did not provide a formal explanation for this robustness. Investigating the theoretical properties underlying this performance would be a natural follow-up.

Our work focused on non-hierarchical data. A logical extension would be to adapt hybrid strategies to multilevel or clustered datasets, which are common in biomedical and epidemiological studies. Additionally, a dedicated R package implementing the hybrid methods with a flexible \(\kappa\) interface would greatly facilitate their use. Such a tool would help analysts explore bias-variance trade-offs across a grid of \(\kappa\) values and select the most appropriate hybrid configuration for their specific data context.

Lastly, promising research directions include incorporating hybrid strategies into Bayesian frameworks, developing adaptive procedures for selecting \(\kappa\) based on predictive metrics, and applying the framework to longitudinal or time-varying data.

\vspace*{3cm}
\noindent {\textbf{Acknowledgement:}}  Abdoulaye Dioni, would like to acknowledge the Canadian Francophonie Scholarship Program and the government of Burkina Faso for their financial support during his PhD studies.

\noindent {\textbf{Conflict of interest statement:}} The authors have declared no conflict of interest.

\noindent{\textbf{Data availability statement :}} The data are accessible via the \texttt{Hmisc} package (version 5.2.3).

\newpage
\appendix

\section*{\Huge{Hybrid methods for missing categorical covariates in Cox model: Supplementary material} }
\pagenumbering{Alph}
\vspace{1cm}

\section{Additional Results from the Simulation Based on the SUPPORT Data}
\label{simulation:resul:chap3}

{
\setlength{\tabcolsep}{10pt}
\begin{longtable}{lcccccc}
\caption{Relative bias (\%) of classical and hybrid methods under a MAR mechanism for \textit{Income} (SUPPORT data), with corresponding $\kappa$ values.}
\label{tab:bias_support}\\
\hline
\rowcolor{gray!40} \textbf{Method} & $Income 11$–$25k$ & $Income >25k$ & Respiratory & Sodium & tt(logAge) & $\kappa$ \\
\hline
\endfirsthead

\hline
\rowcolor{gray!40} \textbf{Method} & $Income 11$–$25k$ & $Income >25k$ & Respiratory & Sodium & tt(logAge) & $\kappa$ \\
\hline
\endhead

\hline
\multicolumn{7}{c}{(Continued on next page)}\\
\endfoot

\hline
\endlastfoot

 \textbf{CC}       & -94.361 & -38.641 & 7.999  & 166.050  & -5.911 & \\
\rowcolor{gray!40} \textbf{IPW}      & 7.435   & 15.887  & -0.171 & -23.163  & 3.148  & \\
 \textbf{MICE(P)}  & -83.952 & -13.922 & 0.095  & 0.349    & 0.094  & \\
\rowcolor{gray!40} \textbf{RF}       & 30.366  & 34.721  & -0.252 & -0.724   & 0.053  & \\

 \textbf{HYBRID 1} & -26.038 & 9.424   & -0.126 & -0.178   & 0.071  & \\

\rowcolor{gray!40} \textbf{HYBRID 2} & -19.518 & 5.909   & -3.177 & -54.691  & 0.502  & 1 \\
 \textbf{HYBRID 3} & 45.275  & 34.055  & -3.504 & -55.280  & 0.514  & 1 \\
\rowcolor{gray!40} \textbf{HYBRID 4} & 13.663  & 19.373  & -3.371 & -54.988  & 0.507  & 1 \\

 \textbf{HYBRID 2} & -27.725 & -1.887  & -1.918 & -15.377  & -1.594 & 0.5 \\
\rowcolor{gray!40} \textbf{HYBRID 3} & -3.458  & 3.088   & -2.055 & -15.919  & -1.497 & 0.5 \\
 \textbf{HYBRID 4} & -12.603 & 1.499   & -2.028 & -15.591  & -1.555 & 0.5 \\

\rowcolor{gray!40} \textbf{HYBRID 2} & -31.112 & -5.018  & -1.355 & -0.752   & -2.440 & 0.3 \\
 \textbf{HYBRID 3} & -17.155 & -6.284  & -1.430 & -1.274   & -2.323 & 0.3 \\
\rowcolor{gray!40} \textbf{HYBRID 4} & -20.637 & -4.368  & -1.443 & -0.930   & -2.393 & 0.3 \\

 \textbf{HYBRID 2} & -35.932 & -9.470  & -0.581 & 19.369   & -3.633 & 0 \\
\rowcolor{gray!40} \textbf{HYBRID 3} & -33.870 & -18.189 & -0.575 & 18.873   & -3.496 & 0 \\
 \textbf{HYBRID 4} & -30.856 & -12.107 & -0.641 & 19.244   & -3.577 & 0 \\
\end{longtable}
}

{
\setlength{\tabcolsep}{10pt}
\begin{longtable}{lcccccc}
\caption{Confidence interval width (95\%) of classical and hybrid methods under a MAR mechanism for \textit{Income} (SUPPORT data), with corresponding $\kappa$ values.}
\label{tab:ciwidth_support}\\
\hline
\rowcolor{gray!40} \textbf{Method} & $Income 11$–$25k$ & $Income >25k$ & Respiratory & Sodium & tt(logAge) & $\kappa$ \\
\hline
\endfirsthead

\hline
\rowcolor{gray!40} \textbf{Method} & $Income 11$–$25k$ & $Income >25k$ & Respiratory & Sodium & tt(logAge) & $\kappa$ \\
\hline
\endhead

\hline
\multicolumn{7}{c}{(Continued on next page)}\\
\endfoot

\hline
\endlastfoot

 \textbf{CC}       & 0.455 & 0.432 & 0.021 & 0.034 & 0.174 & \\
\rowcolor{gray!40} \textbf{IPW}      & 0.458 & 0.431 & 0.022 & 0.035 & 0.178 & \\
 \textbf{MICE(P)}  & 0.428 & 0.408 & 0.017 & 0.026 & 0.142 & \\
\rowcolor{gray!40} \textbf{RF}       & 0.443 & 0.413 & 0.016 & 0.026 & 0.143 & \\

 \textbf{HYBRID 1} & 0.431 & 0.407 & 0.016 & 0.026 & 0.143 & \\

\rowcolor{gray!40} \textbf{HYBRID 2} & 0.431 & 0.409 & 0.017 & 0.027 & 0.145 & 1 \\
 \textbf{HYBRID 3} & 0.450 & 0.418 & 0.017 & 0.027 & 0.145 & 1 \\
\rowcolor{gray!40} \textbf{HYBRID 4} & 0.438 & 0.411 & 0.017 & 0.027 & 0.145 & 1 \\

 \textbf{HYBRID 2} & 0.439 & 0.420 & 0.019 & 0.030 & 0.183 & 0.5 \\
\rowcolor{gray!40} \textbf{HYBRID 3} & 0.458 & 0.429 & 0.019 & 0.030 & 0.184 & 0.5 \\
 \textbf{HYBRID 4} & 0.442 & 0.420 & 0.019 & 0.030 & 0.183 & 0.5 \\

\rowcolor{gray!40} \textbf{HYBRID 2} & 0.444 & 0.426 & 0.020 & 0.032 & 0.204 & 0.3 \\
 \textbf{HYBRID 3} & 0.463 & 0.436 & 0.020 & 0.032 & 0.205 & 0.3 \\
\rowcolor{gray!40} \textbf{HYBRID 4} & 0.445 & 0.426 & 0.020 & 0.032 & 0.204 & 0.3 \\

 \textbf{HYBRID 2} & 0.450 & 0.435 & 0.022 & 0.035 & 0.232 & 0 \\
\rowcolor{gray!40} \textbf{HYBRID 3} & 0.471 & 0.446 & 0.022 & 0.035 & 0.233 & 0 \\
 \textbf{HYBRID 4} & 0.451 & 0.434 & 0.022 & 0.035 & 0.233 & 0 \\
\end{longtable}
}

{
\setlength{\tabcolsep}{10pt}
\begin{longtable}{lcccccc}
\caption{Coverage rate (95\%) of classical and hybrid methods under a MAR mechanism for \textit{Income} (SUPPORT data), with corresponding $\kappa$ values.}
\label{tab:coverage_support}\\
\hline
\rowcolor{gray!40} \textbf{Method} & $Income 11$–$25k$ & $Income >25k$ & Respiratory & Sodium & tt(logAge) & $\kappa$ \\
\hline
\endfirsthead

\hline
\rowcolor{gray!40} \textbf{Method} & $Income 11$–$25k$ & $Income >25k$ & Respiratory & Sodium & tt(logAge) & $\kappa$ \\
\hline
\endhead

\hline
\multicolumn{7}{c}{(Continued on next page)}\\
\endfoot

\hline
\endlastfoot

 \textbf{CC}       & 96.5 & 96.5 & 95.0 & 89.1 & 95.6 & \\
\rowcolor{gray!40} \textbf{IPW}      & 97.7 & 97.0 & 97.1 & 96.7 & 96.3 & \\
 \textbf{MICE(P)}  & 95.9 & 95.9 & 96.5 & 95.2 & 96.6 & \\
\rowcolor{gray!40} \textbf{RF}       & 95.5 & 95.7 & 96.7 & 95.1 & 96.6 & \\

 \textbf{HYBRID 1} & 96.1 & 96.6 & 96.5 & 95.1 & 96.7 & \\

\rowcolor{gray!40} \textbf{HYBRID 2} & 96.0 & 96.2 & 97.3 & 95.9 & 96.1 & 1 \\
 \textbf{HYBRID 3} & 95.6 & 95.5 & 97.3 & 96.0 & 96.1 & 1 \\
\rowcolor{gray!40} \textbf{HYBRID 4} & 95.8 & 95.9 & 97.3 & 96.0 & 96.1 & 1 \\

 \textbf{HYBRID 2} & 98.0 & 98.3 & 95.1 & 95.1 & 93.8 & 0.5 \\
\rowcolor{gray!40} \textbf{HYBRID 3} & 97.5 & 98.5 & 95.3 & 95.2 & 94.0 & 0.5 \\
 \textbf{HYBRID 4} & 97.9 & 98.2 & 95.2 & 95.2 & 94.0 & 0.5 \\

\rowcolor{gray!40} \textbf{HYBRID 2} & 98.8 & 98.9 & 94.0 & 94.4 & 92.7 & 0.3 \\
 \textbf{HYBRID 3} & 98.0 & 99.3 & 94.1 & 94.4 & 92.5 & 0.3 \\
\rowcolor{gray!40} \textbf{HYBRID 4} & 98.5 & 99.0 & 94.2 & 94.3 & 92.6 & 0.3 \\

 \textbf{HYBRID 2} & 99.3 & 99.6 & 93.4 & 93.8 & 90.3 & 0 \\
\rowcolor{gray!40} \textbf{HYBRID 3} & 99.0 & 99.6 & 93.3 & 93.8 & 90.1 & 0 \\
 \textbf{HYBRID 4} & 99.1 & 99.8 & 93.4 & 93.6 & 90.3 & 0 \\
\end{longtable}
}

\section{Kaplan-Meier Curves} \label{courbes:kaplan:meir}

The log-rank test shows no significant survival difference across \textit{income} groups (\(p = 0.57\)).

\begin{figure}[H]  
    \centering
    \includegraphics[width=0.95\textwidth]{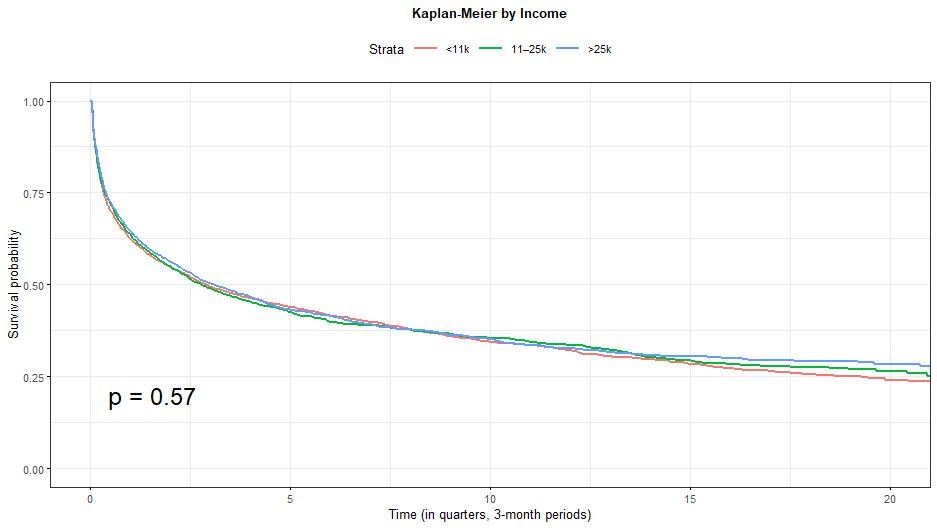}
   \caption{Kaplan-Meier Curves for Income }

   \label{km}
\end{figure}

\section{Checking the assumptions of the Cox model} \label{linearity:PH}

\subsection{Checking linearity Assumption}

If the relationship between the covariates and the log-risk is linear, the martingale residuals should be randomly scattered around zero when plotted against the covariates \cite{Moore2016}. The presence of a non-linear trend suggests a departure from the linearity assumption \cite{Therneau:2000}.
The linearity assumption was evaluated for the continuous covariates: \textit{age at admission} (age), \textit{respiratory rate} (resp), and \textit{serum sodium level} (sod), by testing several common transformations (linear, logarithmic, quadratic, square root, inverse, and natural spline). Model performance was compared using the Akaike Information Criterion (AIC).
For age, the logarithmic transformation yielded the best trade-off, achieving the lowest AIC with minimal model complexity, and was further supported by martingale residual plots. For resp and sodium, the linear form was retained for simplicity, as AIC improvements from alternative transformations were negligible. These modeling choices align with the assumptions used in the simulation data generation process.

\begin{figure}[H]  
    \centering
    \includegraphics[width=0.95\textwidth]{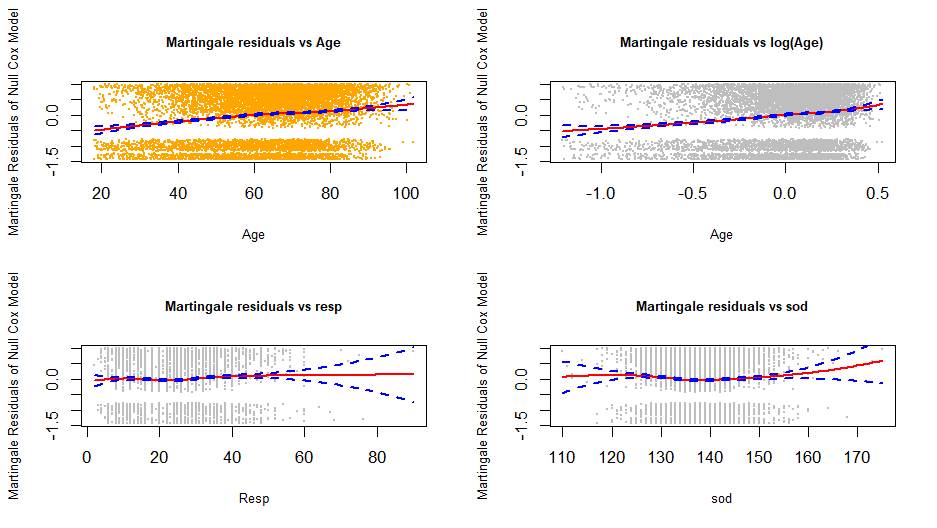}
    \caption{Martingrale residual vs. variable }
\end{figure}

\subsection{Checking proportional hazards assumption}  

The Cox model's proportional hazards assumption stipulates that for two individuals with covariates \( X_i \) and \( X_j \), the ratio of the hazard functions, \( \exp(\beta (X_j - X_i)) \), is independent of time.

\subsubsection*{Graphical approach: Log-Log Survival Curves}

This method is particularly useful when the variable of interest is categorical. Graphically, the proportional hazards (PH) assumption is validated if the log-log curves for different categories remain parallel and do not cross. If they do cross, the PH assumption is violated \cite{Moore2016, Kleinbaum2012}. The plots below show that the assumption is met for the variable (income), while moderate violation is observed for the variable (sex). 

\begin{figure}[H]  
    \centering
    \includegraphics[width=0.85\textwidth]{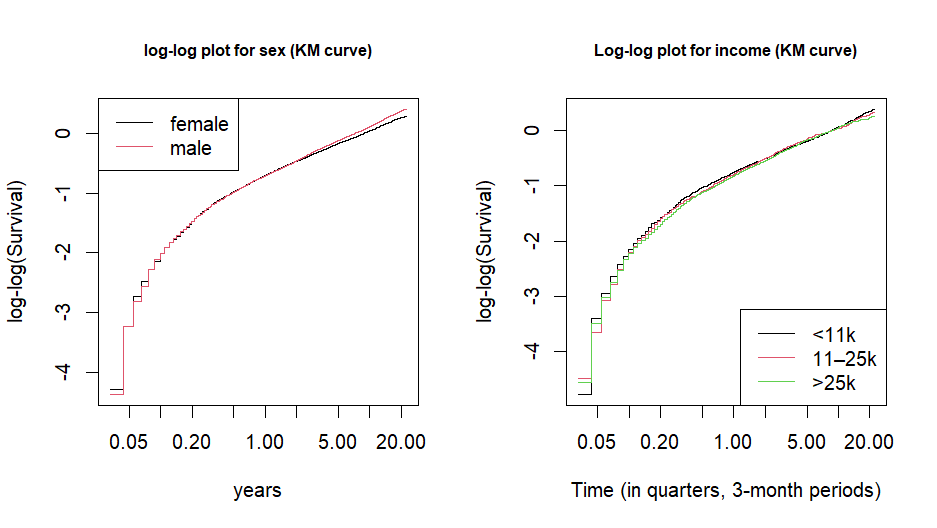}
   \caption{Log-log curve for categorical variables}
   \label{PH_hyp:loglogplot}
\end{figure}

\subsubsection*{Schoenfeld Residuals}  

This approach relies on a visual assessment of the proportional hazards assumption. If this assumption holds, the Schoenfeld residuals should be randomly distributed around zero, without any systematic trend over time. Any marked deviation or apparent trend suggests a violation of this assumption.  
In the plots below, the Schoenfeld residuals visually show no obvious trend, indicating a general adherence to the assumption for these variables. However, further analysis is needed to confirm this observation.

\begin{figure}[H]  
    \centering
    \includegraphics[width=0.95\textwidth]{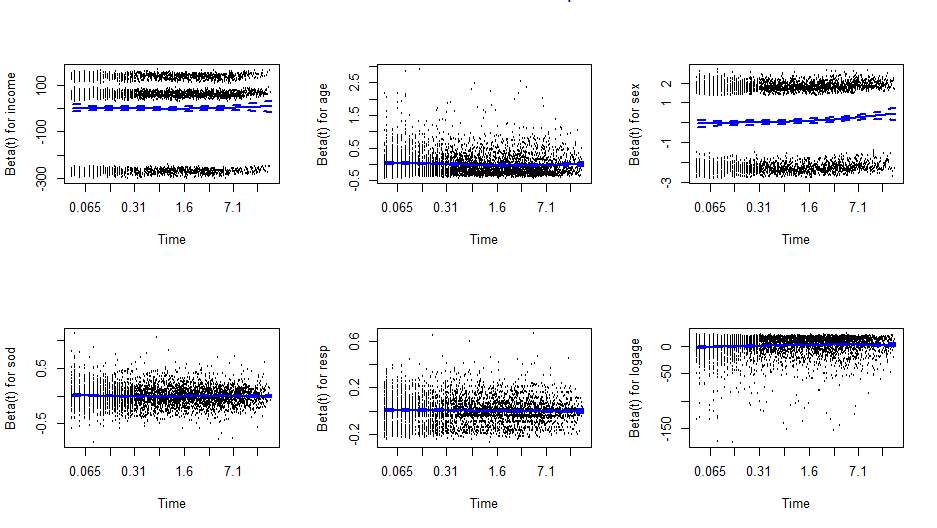}
   \label{PH_hyp:variable2}
   \caption{Schoenfeld Residuals Plots}
\end{figure}

\subsubsection*{Cox Proportional Hazards Assumption Test via \texttt{cox.zph}}

This section reports the results of the proportional hazards (PH) assumption test based on the \texttt{cox.zph} function \cite{Therneau2024}, which evaluates the constancy of covariate effects over time using Schoenfeld residuals. If the assumption holds, no systematic association should be observed between residuals and time.

The results indicate that the proportional hazards (PH) assumption is reasonably satisfied for the variables \textit{income} and \textit{sodium} ($p > 0.05$). However, significant violations are observed for \textit{age} and \textit{sex}, with a moderate violation for \textit{respiratory rate}. The global test is also highly significant ($p < 10^{-9}$), indicating that the PH assumption is not fully met and that appropriate model adjustments are required (see Section~\ref{discussion:chap2}).

\vspace{1em}

\begin{table}[H]
\centering
\caption{Cox Proportional Hazards Assumption Test (Schoenfeld residuals) for the SUPPORT Data}
\label{tab:cox_zph_updated}
\tiny
\begin{adjustbox}{width=0.95\textwidth}
{
\setlength{\tabcolsep}{6pt}
\begin{tabular}{lccc}
\hline
\textbf{Variable} & \textbf{Chi-sq} & \textbf{df} & \textbf{P-value} \\
\hline
Income             & 0.0575   & 2 & 0.97165  \\
Age                & 16.6897  & 1 & $4.4 \times 10^{-5}$ \\
Sex                & 11.6934  & 1 & 0.00063 \\
Sodium             & 1.3110   & 1 & 0.25221 \\
Respiratory rate   & 4.1472   & 1 & 0.04170 \\
\textbf{GLOBAL}    & 56.2447  & 7 & $8.4 \times 10^{-10}$ \\
\hline
\end{tabular}
}
\end{adjustbox}
\end{table}

\end{document}